\documentclass[a4paper,twocolumn,11pt, accepted=2024-03-11]{quantumarticle}
\pdfoutput=1
\usepackage[utf8]{inputenc}
\usepackage[english]{babel}
\usepackage[T1]{fontenc}
\usepackage{amsmath}
\usepackage{diagbox}
\usepackage{amssymb}
\usepackage{soul}
\usepackage{hyperref}
\usepackage{float}
\usepackage{enumitem}
\usepackage{tikz}
\usetikzlibrary{quantikz}
\usepackage{dsfont}
\usepackage{bbold}
\usepackage{cancel}

\usepackage{cite}
\usepackage{todonotes}

\newtheorem{lem}{Lemma}

\newtheorem{defn}{Definition}
\usepackage {ulem}

\usepackage{multirow}

\begin{document}

\author{Javier Gonzalez-Conde}

\affiliation{Department of Physical Chemistry, University of the Basque Country UPV/EHU, Apartado 644, 48080 Bilbao, Spain}
\affiliation{EHU Quantum Center, University of the Basque Country UPV/EHU, Apartado 644, 48080 Bilbao, Spain}
\email{javier.gonzalezc@ehu.eus}

\author{Thomas W. Watts}

\affiliation{School of Applied and Engineering Physics, Cornell University, Ithaca, NY 14853, USA}

\author{Pablo Rodriguez-Grasa}

\affiliation{Department of Physical Chemistry, University of the Basque Country UPV/EHU, Apartado 644, 48080 Bilbao, Spain}
\affiliation{EHU Quantum Center, University of the Basque Country UPV/EHU, Apartado 644, 48080 Bilbao, Spain}
\affiliation{TECNALIA, Basque Research and Technology Alliance (BRTA), 48160 Derio, Spain}

\author{Mikel Sanz}

\affiliation{Department of Physical Chemistry, University of the Basque Country UPV/EHU, Apartado 644, 48080 Bilbao, Spain}
\affiliation{EHU Quantum Center, University of the Basque Country UPV/EHU, Apartado 644, 48080 Bilbao, Spain}
\affiliation{IKERBASQUE, Basque Foundation for Science, Plaza Euskadi 5, 48009, Bilbao, Spain}
\affiliation{Basque Center for Applied Mathematics (BCAM), Alameda de Mazarredo, 14, 48009 Bilbao, Spain}

\title{Efficient quantum amplitude encoding of polynomial functions}

\begin{abstract}
Loading functions into quantum computers represents an essential step in several quantum algorithms, such as quantum partial differential equation solvers. Therefore, the inefficiency of this process leads to a major bottleneck for the application of these algorithms. Here, we present and compare two efficient methods for the amplitude encoding of real polynomial functions on $n$ qubits. This case holds special relevance, as any continuous function on a closed interval can be uniformly approximated with arbitrary precision by a polynomial function. The first approach relies on the matrix product state representation (MPS). We study and benchmark the approximations of the target state when the bond dimension is assumed to be small. The second algorithm combines two subroutines. Initially we encode the linear function into the quantum registers either via its MPS or with a shallow sequence of multi-controlled gates that loads the linear function's Hadamard-Walsh series, and we explore how truncating the Hadamard-Walsh series of the linear function affects the final fidelity. Applying the inverse discrete Hadamard-Walsh transform converts the state encoding the series coefficients into an amplitude encoding of the linear function. Thus, we use this construction as a building block to achieve an exact block encoding of the amplitudes corresponding to the linear function on $k_0$ qubits and apply the quantum singular value transformation that implements a polynomial transformation to the block encoding of the amplitudes. This unitary together with the Amplitude Amplification algorithm will enable us to prepare the quantum state that encodes the polynomial function on $k_0$ qubits. Finally we pad $n-k_0$ qubits to generate an approximated encoding of the polynomial on $n$ qubits, analyzing the error depending on $k_0$. In this regard, our methodology proposes a method to improve the state-of-the-art complexity by introducing controllable errors.

\end{abstract}

\maketitle

\section{Introduction}
Over the past few decades, there has been a significant interest in quantum computing due to its theoretical capacity to surpass classical information processing for certain specific application areas. Despite the fact that current quantum computers are hindered by noise and decoherence, there have been successful experimental demonstrations of quantum advantage \cite{google,Wu_2021, Zhong_2020} and even recently the first logical quantum processor was announced \cite{lisq}. However, these achievements have yet to have any practical relevance, leaving the search for useful applications ongoing.
Many promising quantum algorithms, such as solving systems of linear equations \cite{HHL, HHL_Childs}, performing data fitting \cite{Data_Fitting}, computing scattering cross sections \cite{Preconditioned_HHL, Scherer_2017}, pricing financial derivatives \cite{Patrick_Options, Stamatopoulos_2020,Martin_2021, gonzalezconde2022simulating, finance, ORUS2019100028, Egger_2020, agliardi2023quadratic}, and in general solving differential equations \cite{leyton2008quantum, Berry_2017, Liu_2021, Zanger_2021, Garc_a_Ripoll_2021, Pablo, lin2022koopman,JIN2023112149, lewis2023limitations, an2022efficient, PhysRevResearch.2.043102, Sornborger2023, Sornborger2023_2, PhysRevA.105.052404,Gay_Balmaz_2022, MixedSC, bondar2019koopman}, require the efficient loading of classical data into quantum devices. Unfortunately, this step remains a challenging problem, and it is a major bottleneck for the practical application of quantum computation, especially within the emerging field of quantum machine learning in the NISQ-Era \cite{Preskill_2018,Havl_ek_2019,rigorous_robust}.

In this regard, the main drawback comes from the fact there is no universal loading protocol and each particular case must be carefully studied in order to design bespoke encoding protocols that cater to the specific problem to be solved \cite{Encoding_variational, Schuld2021, lloyd2020quantum}. In this sense, one of the main embedding techniques is the amplitude encoding, which loads the values of a discretized, normalized complex function into the amplitude of the quantum states \cite{Li_2023, mcardle2022quantum, mottonen2004transformation, sun2023asymptotically, PhysRevResearch.3.043200, Araujo_2021, zhao2019state, PhysRevLett.85.1334, PhysRevLett.122.020502, Bausch2022fastblackboxquantum, grover2002creating, rattew2022preparing, Wang_2022, PhysRevLett.129.230504, Marin_Sanchez_2023, PhysRevResearch.4.023136, zoufal2019, zylberman2023efficient, Moosa_2023, Grasedyck2010PolynomialAI,holmes2020efficient, holmes2020entanglement, Melnikov_2023}.
Several approaches have already been presented in the literature for implementing the amplitude embedding, many requiring a huge (exponential) number of resources (ancillas and gates) \cite{mottonen2004transformation, sun2023asymptotically, PhysRevResearch.3.043200, Araujo_2021, zhao2019state},  oracles \cite{PhysRevLett.85.1334, PhysRevLett.122.020502, Bausch2022fastblackboxquantum, grover2002creating, rattew2022preparing, Wang_2022}, sparsity in the quantum state \cite{PhysRevLett.129.230504}, training variational circuits \cite{Marin_Sanchez_2023, PhysRevResearch.4.023136, zoufal2019}, truncating a series expansion \cite{zylberman2023efficient, Moosa_2023}, use the quantum singular value transformation \cite{mcardle2022quantum, Li_2023}   or matrix product states \cite{Grasedyck2010PolynomialAI,holmes2020efficient, holmes2020entanglement, Melnikov_2023, PhysRevResearch.4.043007, Lin_2021, LUBASCH2018587, PhysRevA.101.010301, Gourianov_2022, iaconis2023quantum}.

There exists a necessity for loading polynomial functions due to the growing number of applications of quantum computing. Specifically, in finance  \cite{Patrick_Options, Stamatopoulos_2020,Martin_2021, gonzalezconde2022simulating, finance, ORUS2019100028, Egger_2020, agliardi2023quadratic}, the ability to efficiently load first order polynomials, $f(j) =~ aj + b$ allows for options pricing via quantum amplitude estimation (QAE) without coherent arithmetic \cite{generalizedinnerproduct}. In this sense, applying QAE allows us to extract the amplitude $\sum_j f(j)p(j) = \mathbb{E}[f(X)]$ where $X$ is a random variable with probability mass function $p(j)$. This approach can be generalized to the multidimensional setting where the ability to load multivariate linear functions yields efficient algorithms for pricing basket options \cite{optionspricingwithqc}. Furthermore, quantum circuits for efficiently loading the linear function can be used to construct a block encoding of the identity function thus allowing us to apply the quantum singular value Transformation \cite{PhysRevX.6.041067, PhysRevLett.118.010501, Low2019hamiltonian, Gily_n_2019, tang2023cs, PhysRevA.103.042419} (QSVT) in order to obtain any polynomial amplitude encoding \cite{mcardle2022quantum, Guo_2021, rattew2023nonlinear, Li_2023}. This is a powerful tool capable of uniformly approximating the encoding of any continuous real function defined on a closed interval with arbitrary precision \cite{Polynomials1, Polynomials2 }.

 In this article we present two methods for implementing the amplitude encoding of real valued polynomials into quantum computers with linear complexity. The first one is based on the matrix product sate (MPS) representation of quantum states and its implementation on a quantum computer. We explore how approximating the MPS affects the achieved fidelity and the resources requirements \cite{roman_review, Vidal_2003, Verstraete_2008, perezgarcia2007matrix, Ran_2020, malz2023preparation}.   On the other hand, the second method consists of two steps, first we propose a novel protocol to efficiently load the linear function on $k_0$ qubits based of the discrete Hadamard-Walsh transform  (DHWT) \cite{walsh} that we use to achieve a block encoding of the amplitudes with complexity $\mathcal{O}(k_0)$.  Secondly, we use the QSVT to implement a polynomial transformation on the eigenvalues of the block encoding to achieve the desired target state \cite{Guo_2021, rattew2023nonlinear}. Note that as we have implemented a block encoding of the linear function, our method avoids errors due to the polynomial approximation of the $\text{arcsin}(x)$ as proposed in previous works \cite{mcardle2022quantum,Li_2023}. Furthermore, we reduce the complexity of the state-of-the-art for encoding polynomials functions via QSVT by introducing a controllable error.

The article is structured as follows, first we review the loading of polynomials via MPS. Next, we introduce our new methodology that combines DHWT to load the linear function with the QSVT to implement the polynomial transformation of the amplitudes. Finally we show numerical results and compare our method with previous results in the literature.

\section{Loading of polynomials via MPS}
In this section we analyze the methods based on MPSs to encode polynomials into the amplitude of a quantum state according to following definition.

\begin{defn} Let $P(x): [a,b] \to \mathbb{C}$, be a polynomial with complex coefficients. We define the $n$-qubit normalized representative state of $P(x)$ as the quantum state $\ket{\Phi_P}=~\frac{1}{C_P}\sum_{j=0}^{2^n-1}P(x_j)\ket{j}$, with $x_j=~a+~j\frac{b-a}{2^n-1}$ and $C_P$ the normalization factor. 
\label{def1}
\end{defn}
In the particular case of the linear function, i.e. $P(x)=~x$, defined on $[0, 2^n-1]$, we have the normalization factor is $C_P=~\sqrt{(2^{n+1}-1)2^n/(6(2^n-1))}$. For now on for simplicity, we will assume that the polynomials we work with are defined on the interval $[0,1]$, so that might include rescaling of the domain of the linear function.\\

The complete description of a quantum state of $n$ linearly connected qubits (sites) can be represented by a tensor, $A$, with $n$ physical indices. A state of this kind is referred to as a matrix product state (MPS) \cite{roman_review, Vidal_2003, Verstraete_2008, perezgarcia2007matrix}. Each physical index is assigned to a qubit and has a degree of $d = 2$ i.e., the index is either $0$ or $1$. For a specific choice of each physical index $j_i$, the tensor's values give rise to a collection of $n$ matrices whose product is equal to the amplitudes of the computational basis state $\ket{j_{n-1}\ldots j_0}$. For open boundary conditions, the MPS representation of a quantum state is given by
\begin{equation*}
    |\Psi\rangle = \sum_{j_0\cdots j_{n-1}}\sum_{\alpha_1\cdots \alpha_{n-1}} A_{j_0,\alpha_1}^{[1]} A_{j_1,\alpha_1\alpha_2}^{[2]} \\
\end{equation*}
     
\begin{equation}
  \ \ \ \  \cdots A_{j_{n-1},\alpha_{n-1}}^{[n]} \ket{j_{n-1}\ldots j_0}.
\end{equation}

This representation has $n-2$ tensors of order 3, denoted as $A^{[i]}_{j_{i-1},\alpha_{i-1}\alpha_{i}}$, $\forall \ i \neq 0,n-1$, and 2 external tensors $A_{j_0,\alpha_1}^{[1]}$ and $A_{j_{n-1},\alpha_{n-1}}^{[n]}$ of order 2, with $j_i$  the physical indexes that range from 0 to $d-1$ and $\alpha_i$ the virtual indices from 0 to $\chi_i-1$. The virtual dimensions, $\chi_i$, connecting each pair of tensors via the virtual indices are referred as the bond dimensions. We define the bond dimension of the entire MPS as $\chi=\text{max}_i\ \chi_i$.

When it comes to representing polynomials as quantum states, Grasedyck \cite{Grasedyck2010PolynomialAI} proved that for a real valued polynomial of degree $d$ encoded in the amplitudes of a quantum state according to Def. \ref{def1}, the MPS bond dimension is as much $\chi =~ d +1$.

\subsection{Obtaining the exact MPS}
\begin{figure}[t!]
\centering
\includegraphics[width=1\columnwidth]{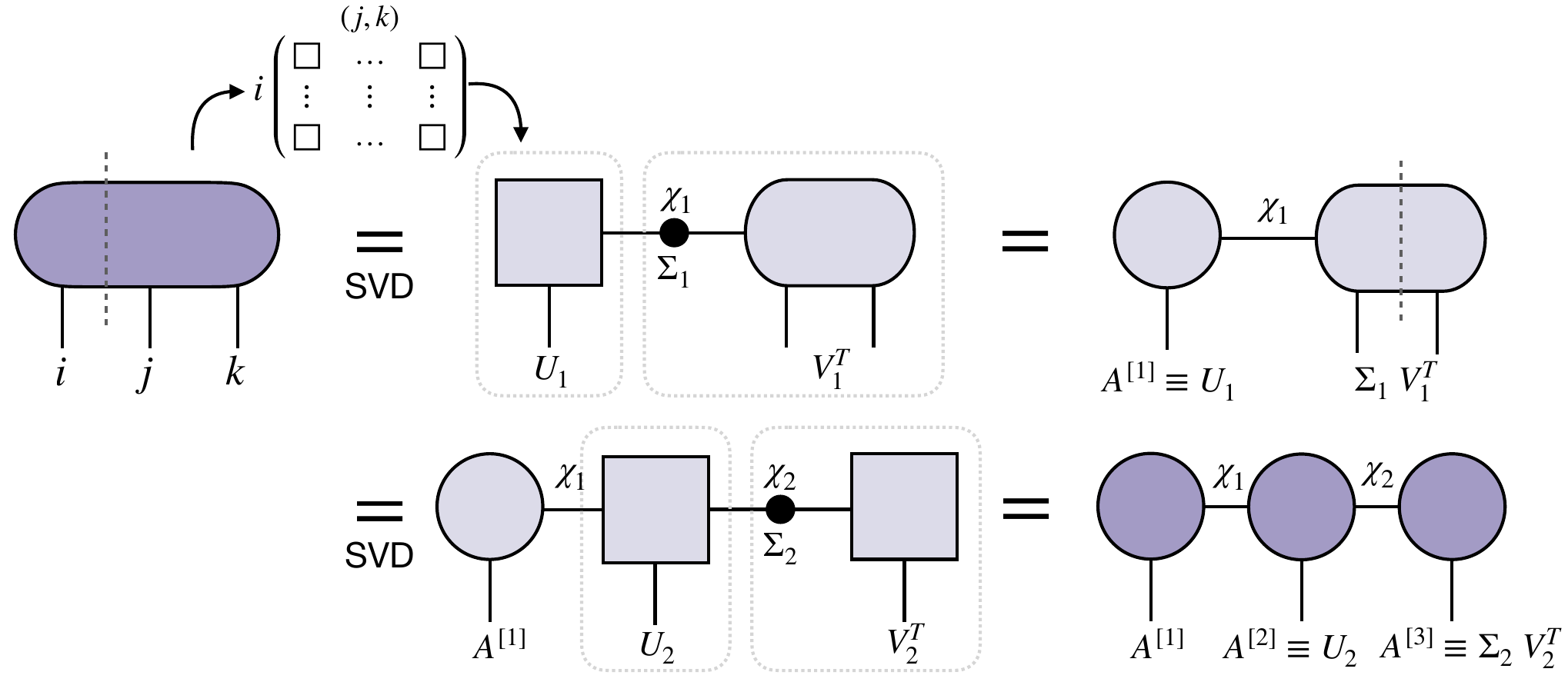}
\caption{Iterative  singular value decomposition (SVD) \cite{Wall2003} procedure for obtaining the MPS from a tensor with $n=3$ physical indices. The process involves $n-1$ uses of the SVD. Notably, the matrices containing the singular values are absorbed to the right, resulting in the left canonical form of the MPS.}
\label{Fig:Tensor_to_MPS}
\end{figure}

The MPS representation of a quantum state $|\Psi\rangle$ is not unique, as different choices of $A_{j_i,\alpha_{i-1}\alpha_i}^{[i]}$ can yield the same quantum state. We focus on the left canonical form, which implies the following conditions
\begin{align}
    &\sum_{j_0,\alpha_1} A_{j_0,\alpha_1}^{[1]} A_{j_0,\alpha_1}^{[1]\dagger}=1,\\
    &\sum_{j_i,\alpha_i} A_{j_i,\alpha_{i-1}\alpha_i}^{[i]} A_{j_i,\alpha'_{i-1}\alpha_i}^{[i]\dagger}=\delta_{\alpha_{i-1}\alpha'_{i-1}},\\
    &\sum_{j_{n-1}} A_{j_{n-1},\alpha_{n-1}}^{[n]} A_{j_{n-1},\alpha'_{n-1}}^{[n]\dagger}=\delta_{\alpha_{n-1}\alpha'_{n-1}}.
\end{align}
To obtain the MPS representation of a quantum state, the singular value decomposition (SVD) is employed \cite{Wall2003}. Initially, the quantum state, represented as a tensor $A$ of rank $n$ and dimension 2, is reshaped into a matrix by combining all the indices except one. The SVD is then applied to this matrix, decomposing it into the matrix of left singular vectors, $U$, the matrix of singular values, $\Sigma$, and the matrix of right singular vectors, $V^\dagger$. As we will depict in the following section, it is possible to truncate the smallest singular values by choosing a desired bond dimension $\chi$ and keeping only the $\chi$ largest values. Then, the matrix of singular values is contracted with the matrix of right singular vectors (left canonical form),  and the resulting matrix, $\Sigma V^\dagger$, is reshaped back into a tensor that now has an extra virtual index. This process, shown in Fig. \ref{Fig:Tensor_to_MPS}, is iterated for each physical index. Finally, we obtain an MPS that approximates the original quantum state while providing a compact and efficient representation. 

The computational cost of performing SVD on an $m\times n$ matrix is typically $\mathcal{O}\bigl(\mathrm{min}(m n^2,m^2 n )\bigr)$ and must be taken into account as a preprossessing cost in this methodology. This cost can be reduced for sparse or structured low-rank matrices. In the case of an exact matrix product state (MPS), the bond dimension doubles for each connection between core tensors. This leads to a maximum bond dimension of $2^{\lfloor n/2 \rfloor}$, occurring in the middle of the MPS. As a result, the computational cost of the entire algorithm is primarily determined by the SVD of this central square matrix, i.e.  $\mathcal{O}(2^{3n/2})$, which exhibits exponential scaling with the number of qubits. Therefore, this non-negligible  classical pre-processing cost must be considered in the overall complexity of the MPS algorithm.

In the particular case of the linear function, the analytical expression of the exact MPS \cite{Oseledets}, which has bond dimension $\chi=2$, reads

\begin{equation*}\label{MPS_linear}
   |\Phi_L\rangle=\sum_{j_0\cdots j_{n-1}}
   \begin{pmatrix}
j_0/C & 1\\
\end{pmatrix}
\begin{pmatrix}
1 & 0\\
2 j_1/C & 1
\end{pmatrix}
\end{equation*}
\begin{equation}
\ \ \ \ \ \cdots \begin{pmatrix}
1 \\
 2^{n-1} j_{n}/C
\end{pmatrix}\ket{j_{n-1}\ldots j_0}.
\end{equation}

\subsection{Approximation of the exact MPS}

While, in the worst case scenario, it is possible to exactly represent any quantum state as an MPS by allowing the bond dimensions to grow up to $2^{\lfloor n/2 \rfloor}$, we can potentially achieve an exponential compression by approximating the initial tensor using $\mathcal{O}(2n\chi^2)$ elements, given a fixed bond dimension, $\chi_i = \chi$ $\forall\ i$. However, as the entanglement of the state to be approximated increases, the minimum bond dimension $\chi$ required to get a good description of the state using an approximated MPS representation also grows \cite{schuch2008entropy, schollwock2011density}. Notice that when truncating the maximum bond dimension of the MPS, the complexity of the classical preprocessing becomes $\mathcal{O}(n\chi^{3})$.

In order to estimate the error incurred when truncating the bond dimension, it is necessary to consider that during each iteration of the compression protocol we perform a SVD and discard singular values. The error incurred when approximating a matrix by considering its $k$ largest singular values is determined by the Eckart-Young theorem \cite{eckart1936approximation}, and corresponds to the sum of the omitted singular values. During compression, this rank-$k$ approximation is performed for each core tensor. Thus, the overall error of the MPS approximation can be upper bounded in the Frobenius norm as 
\begin{equation}
\label{errorMPS}
    \lVert A - \tilde{A} \rVert^2_F\leq \sum_{i=1}^{n-1}\left(\sum_{k=\chi_i+1}^{\text{dim}(\Sigma_i)} \sigma_k^2 (\Sigma_i) \right), 
\end{equation}
where $\tilde{A}$ denotes the approximated MPS. This equation encompasses the error contributions from the $n-1$ singular value matrices $\Sigma_i$, each characterized by a fixed bond dimension, $\chi_i$. In order to keep the approximation error low, it is crucial that the spectrum of each $\Sigma_i$ decays rapidly.

In this regard, the state-of-the-art technique for preparing smooth, differentiable, real-valued functions using matrix product states relies on singular values exhibiting exponential decay as demonstrated  in Ref.~\cite{holmes2020efficient}. This allows for good approximations of such a class of functions while maintaining low bond dimension values, thus significantly reducing the required resources to prepare the associated function-encoding quantum state. The method relies on the fact that, for such functions, the entanglement entropy, quantified by the von Neumann entropy, scales logarithmically with the number of qubits, $n$, and therefore, these functions can be efficiently approximated by an MPS with a low bond dimension, as argued in Ref.~\cite{schuch2008entropy}. Although empirical results with this technique applied to polynomials show good performance with $\chi = 2$, the upper bound of the von Neumann entropy depends on the maximum derivative value of the function within the considered interval. Consequently, as we discuss later in this work, for certain polynomials the truncation to $\chi = 2$ does not yield a satisfactory approximation.

In the particular case of the linear function, the analytical expression from Eq.~\ref{MPS_linear} reveals that achieving the exact MPS representation requires a bond dimension of $\chi=2$. However, one might consider reducing the bond dimension to 1 to investigate how severely accuracy decays. We explore this possibility by comparing the linear function approximated by MPSs of bond dimension $\chi=2$ (exact MPS) and $\chi=1$ (product state) in Section \ref{num_results}.

\subsection{From MPS to circuit}
\label{From MPS to circuit}
Let us now analyze the resources needed to translate an MPS in either cases, exact or approximated, into a quantum circuit \cite{Ran_2020, rudolph2022decomposition, perezgarcia2007matrix, PhysRevLett.95.110503}. First we will assume that the bond dimensions are powers of two, padding the tensors with zeros if needed. Therefore, we can assume that these tensors are isometries of $2\chi_i\times \chi_{i+1}$ and thereby can be embedded into a unitary gate acting on $n_{\chi_i}=\text{max}\{\log_2(2\chi_i),\log_2(\chi_{i+1})\}$ qubits. The factor of two arises from the two possible values that each of the physical indices can take. The arrangement of gates in the MPS conversion process, following a linear topology, results in the formation of a single layer of multi-qubit unitaries, whose sizes $n_{\chi_i}$ depends on the associated bond dimensions. These unitaries are organized in a staircase topology, commonly referred to as a linear circuit layer \cite{rudolph2022decomposition}.
Therefore the complexity is equivalent to implementing a cascade of $n-1$ multi-qubit unitaries. In the case $\chi=2$, this complexity scales as $\mathcal{O}(n)$ two-qubit unitaries.
On top of this, the cost of decomposing these unitaries into two-qubit gates must be taken into account. This cost is considerably larger when decomposing multi-qubit unitaries and might result in an exponential number of two-qubit gates. \cite{Shende_2004, Barenco_1995}. Additionally, one might consider implementing circuits that approximate the exact multi-qubit unitaries \cite{rudolph2022decomposition}. Lastly, in Ref.~\cite{malz2023preparation} authors proposed a method for loading translation-invariant short-correlation MPS with an error $\epsilon$ in depth $\mathcal{O}(\log(N/\epsilon))$ thereby establishing a class of MPSs that admit efficient circuit implementations.

We can conclude that even though MPSs encoding polynomials admit an analytical construction, we have to consider the cost of obtaining the SVD and the cost of the decomposition of the associated unitaries. These costs can be significantly reduced by truncating down to a bond dimension of $\chi= 2$, although this introduces an uncontrollable source of error \cite{Schollw_ck_2011, mcardle2022quantum, holmes2020efficient, holmes2020entanglement}. Additionally, one can also consider the possibility of approximating the multi-qubit unitaries of a MPS's circuit representation \cite{rudolph2022decomposition}, which could result in high fidelity states in some cases although there is not a priori guarantee of success.

\section{Efficient loading of polynomials via DHWT and QSVT}
In this section we present a method for loading polynomials by combining a technique to encode the linear function into the amplitudes of a quantum state via the discrete Hadamard-Walsh transform (DHWT) with the quantum Singular Value Transformation (QSVT) algorithm. The first methodology encodes the linear function  with a shallow sequence of multi-controlled gates that loads its Hadamard-Walsh series expansion, followed by the inverse discrete Hadamard-Walsh transform. By truncating the series expansion, we demonstrate that this approach allows for a controllable approximation of the state representing the linear function up to a certain error. We analyze this error in terms of infidelity, $\epsilon$ with respect to the exact state and through the deviation $\delta^{(2)}_j$ in each amplitude from the exact amplitudes.  The second step leverages the results of \cite{Guo_2021, rattew2023nonlinear} to generate a block encoding of the linear function. This involves using the unitary to load the linear function into amplitudes of a $k_0$-qubit quantum state. Then, the Quantum Singular Value Transformation (QSVT) is applied to implement a polynomial transformation, $P(x)$, on the amplitudes of the quantum state corresponding to the identity function. This results into a unitary block encoding of the polynomial that applied to an adequate initial state yields an encoding of the desired polynomial.
Finally, we pad the remaining $n-k_0$ qubits, obtaining an approximated encoding.

\subsection{The Discrete Hadamard-Walsh
transform}

\begin{figure}[t!]
\centering
\includegraphics[width=1\columnwidth]{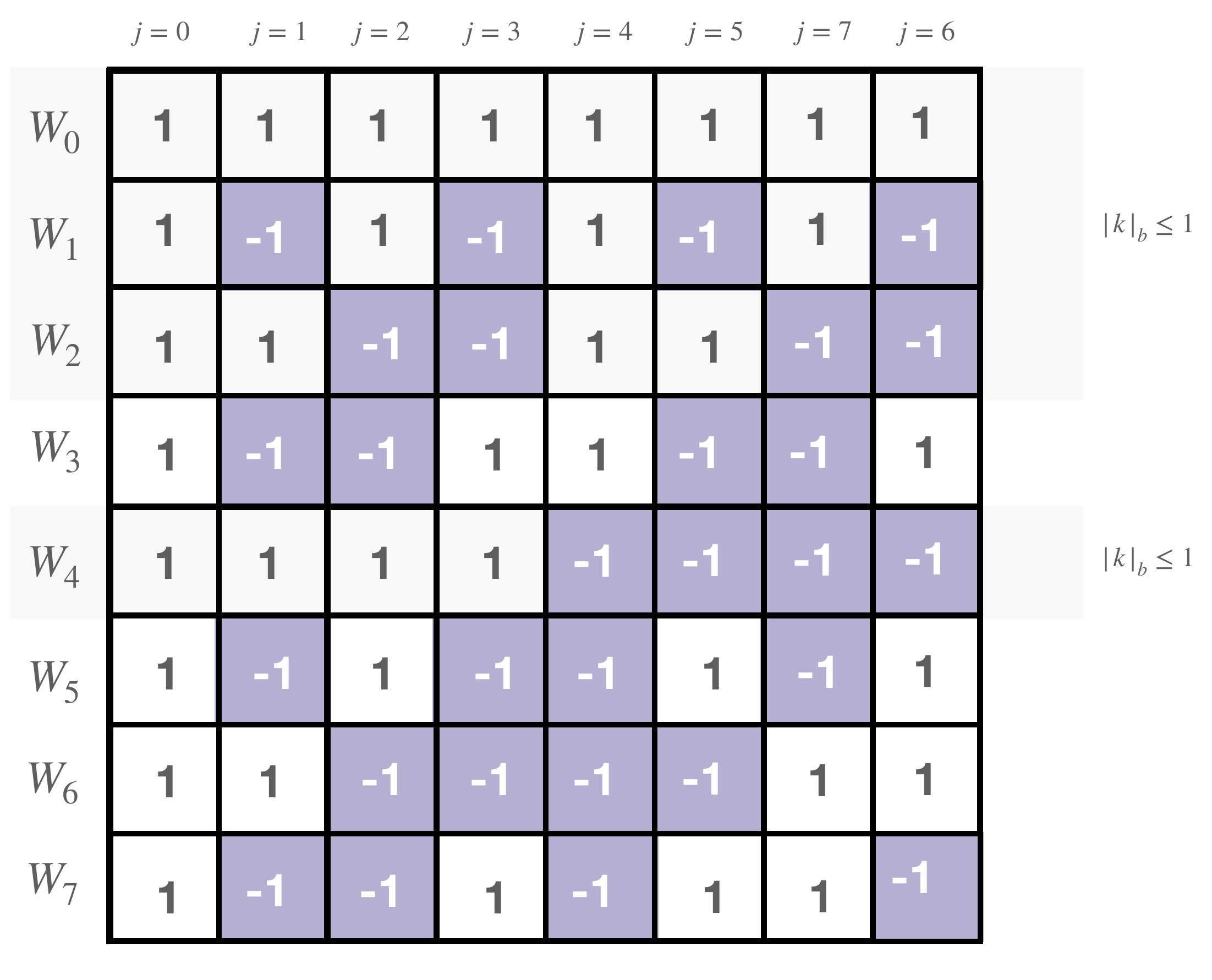}
\caption{Discrete Hadamard-Walsh Transform for $n=~3$ qubits in the natural order representation.}
\label{Fig:circuit_linear}
\end{figure}

\begin{defn}
\label{DHWT}
The discrete Hadamard-Walsh transform (DHWT) is a linear, orthogonal and symmetric operation that transforms discrete signals or sequence of sorted data into a new representation given by the Hadamard-Walsh Series
\begin{equation}
    HWT: (z^{(0)}_n \ldots z^{(N-1)}_{n}) \rightarrow (x^{(0)}_n\ldots x^{(N-1)}_{n}),
\end{equation}
\begin{equation}
    x^{(k)}_n=\frac{1}{\sqrt{N}}\sum_{j=0}^{N-1}z^{(j)}_n W_k(j) 
\end{equation}
where $j=~\sum_{m=0}^{n-1}j_m 2^{m}$, $k=\sum_{m=0}^{n-1}k_m 2^m $ with  $N=2^n$, $j_m, k_m \in \{0,1\}$ and $W_k(j)=~(-1)^{\sum_{m=0}^{n-1} j_m k_m}$ is the $k$-th Walsh function, where we have used the natural order.
\end{defn}
We also define the binary norm of an integer as $|k|_b=~\sum_{m=0}^{n-1} k_m$. Note that when $|k|_b=1$, it is equivalent to say that $k$ is a power of 2. 
Additionally, when representing a quantum state $\ket{j}$ in terms of a binary notation we will denote the state as $\ket{j_{n-1}\ \dots \ j_0}$, taking the order of the tensor product from right to left.

\begin{lem}
\label{lemma_WHT}
    Let $(0,1  \ldots, 2^n-1)$ be the discrete sorted input sequence. Then, the coefficients of its Hadamard-Walsh series are given by 
\begin{equation}
x^{(k)}_n = 
     \begin{cases}
      2^{n/2}(2^n-1)/2  &\quad\text{if} \ k=0\\
       -2^{n/2}k/2 &\quad\text{if} \ |k|_b=1 \\
       0 &\quad\text{otherwise}  \\
     \end{cases}
\end{equation}  
\end{lem} 

Note that in general, the sparsity of the state encoding the Hadamard-Walsh series of a polynomial of degree $d$ represented in $n$ qubits, with $d\leq n$, is $s=\sum_{k=0}^d \binom{n}{k}$ \cite{Welch_2014}. Therefore, one might consider the techniques in Ref. \cite{PhysRevLett.129.230504} to implement the state with depth $\mathcal{O}(\log(ns))$ and  $\mathcal{O}(ns\log(s))$ ancillary qubits.

\subsection{Exact loading of the linear function via DHWT}

In the particular case of the linear function, the target state is  $\ket{\Phi_L}_n=~\frac{1}{C_n}\sum_{j=0}^{2^n-1}j\ket{j}$, with normalization constant $C_n~=~\sqrt{(2^{n+1}-1)(2^n-1)2^n/6}$. 
Notice that Lemma \ref{lemma_WHT} provides the state that encodes the discrete Hadamard-Walsh transform of the coefficients of $\ket{\Phi_L}_n$ which we can write as
$$\ket{\Tilde{\Phi}_{L}}_n=\frac{1}{\tilde{C}_n} \sum_{|k|_b\leq 1} x^{(k)}_n\ket{k},$$
with $\tilde{C}_n$ the corresponding normalization factor. Note that the state above is a $\ket{1}$-sparse quantum state, i.e. the bit string representations  of the basis states whose superposition conforms the state have at most one $\ket{1}$, i.e. $\{ \ket{0 0\ \dots 0 0},\ \ket{1 0\ \ldots 0} \dots, \ket{0 0\ \ldots 0 1}\}.$

Due to its structure, the state $\ket{\Tilde{\Phi}_{L}}_n$ can be efficiently encoded into a quantum computer  with gate complexity $\mathcal{O}(n)$ according to the circuit depicted in Fig. \ref{Fig:circuit_linear}, where the angle of every multi-controlled rotation is given by $\theta_k=~\arcsin \left(2 x^{(2^{k})}_{n} / \tilde{C}_nG_n(k) \right)$, with $k=0,\ \dots , \ n-1$ and $G_n(k)=~\prod_{i=k+1}^{n-1} \cos(\theta_i/2)$ if $k<n-1$ and $G_n(k)=1$ if $k=n-1$. We denote the unitary corresponding to this circuit as $U_{L,n}$. Once that we have encoded the state $\ket{\Tilde{\Phi}_{L}}_n$ that represents the discrete Hadamard-Walsh series of our target state, we can simply uncompute the Hadamard-Walsh transform to obtain $\ket{\Phi_L}_n$. In terms of gates on a quantum computer, this operation is a parallel implementation of Hadamard gates on all the qubits.

Additionally, we would like to remark that due to normalization, any linear function with an arbitrary slope and 0 offset will lead to the same encoding. However, it is possible to vary the slope by changing the offset of the linear function under the restriction of the quantum state to be normalized.

\begin{figure}[t!]
\centering
\includegraphics[width=1.03\columnwidth]{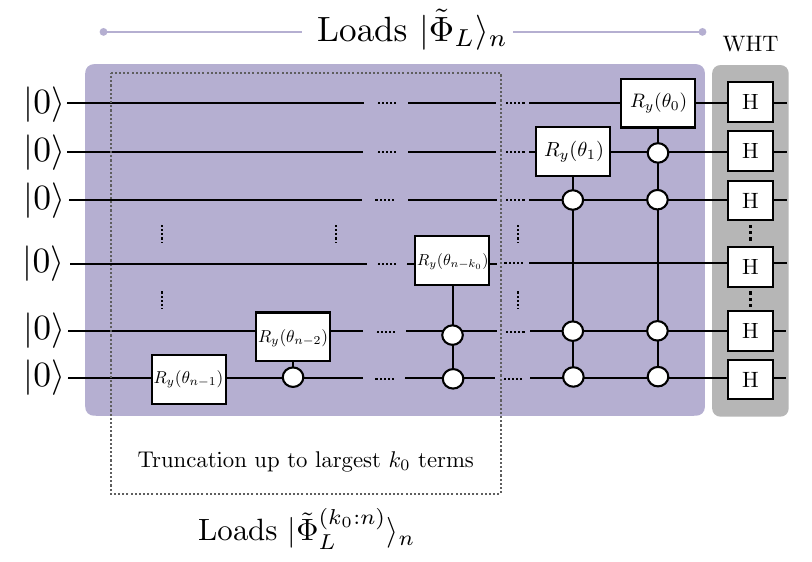}

\caption{Circuit implementation for loading the quantum state $\ket{\Phi_L}_n$, denoted as $U_{L,n}$ with complexity $\mathcal{O}(n)$ multi-controlled gates. The angle of every multi-controlled rotation is given by $\theta_k=~\arcsin \left(2 x_{n}^{(2^{k})} / \tilde{C}_nG_n(k) \right)$, with $k=0\ \dots\ n-1$ and $G_n(k)=~\prod_{i=k+1}^{n-1} \cos(\theta_i/2)$ if $k<n-1$ and $G_n(k)=1$ if $k=n-1$.  The first part of the circuit loads the state encoding its Hadamard-Walsh series $\ket{\Tilde{\Phi}_L}_n$ and once it has been loaded, we apply the Hadamard-Walsh transform to achieve $\ket{\Phi_L}_n$. If we consider only the first $k_0$ rotations, then we get the approximated Hadamard-Walsh series $\ket{\Tilde{\Phi}^{(k_0:n)}_{L}}_n$ and the respectively approximated state $\ket{\Phi^{(k_0:n)}_{L}}_n$.  With this truncation the angles change its value to $\theta^{k_0}_k=~\arcsin (2 x_{n}^{(2^k)}/\tilde{C}^{(k_0:n)}_{n}G^{(k_0:n)}_n(k))$, with $\tilde{C}^{(k_0:n)}_{n}$ the new normalization factor and $G^{(k_0:n)}_n(k)=~\prod_{i=k+1}^{n-1} \cos(\theta^{k_0}_i/2)$ if $n-k_0\leq k<n-1$ and $G^{(k_0:n)}_n(k)=1$ if $k=n-1$ and the complexity is reduced to $\mathcal{O}(k_0)$. We denote this last circuit that loads the approximated state as $U_{L,n}^{(k_0:n)}$. See the appendix in Ref. \cite{Marin_Sanchez_2023} for the details of the decomposition of multi-controlled gates. }
\label{Fig:circuit_linear}
\end{figure}

\begin{figure*} 
\centering
\includegraphics[width=2.1\columnwidth]{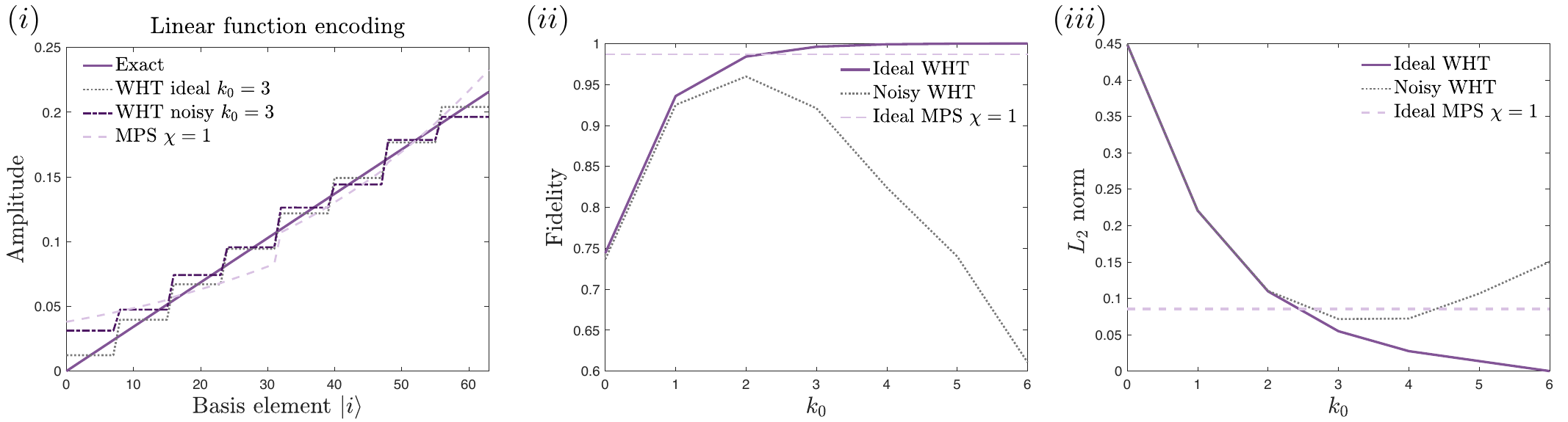}
\caption{Comparison of different methods for loading the linear function on $n=6$ qubits. $(i)$ Illustrates the resulting state by using different encoding protocols, DHWT with $k_0=3$ and MPS $\chi=1$, in noisy an ideal scenarios. The choice of $k_0=3$ is due to the fact this case has a high fidelity and the minimum error measured in the $L_2$ norm for DHWT method in presence of noise. This is shown in $(ii)$ and $(iii)$, where fidelity and error measured in the $L_2$ norm for both the ideal and noisy DHWT methods are plotted with respect to $k_0$. We have not considered noise for the case MPS $\chi=1$ as its loading circuit is only a layer of single qubit rotations, and its impact is marginal.}
\label{linear_analysis}
\end{figure*}

\subsection{Approximated loading of linear polynomials via DHWT}
 
In the preceding section, we have conclusively demonstrated the efficient loading of linear functions. Now, the inevitable query arises: can the Hadamard-Walsh series be deliberately truncated while maintaining control over the process? In other words, is it possible to strike a balance between the number of terms truncated and the resulting error, thereby achieving a quantum state approximation that accurately encodes our intended target?

We now proceed to illustrate how the loading of the linear function can be approximated by truncating the Hadamard-Walsh series. Let us assume that we have the DHWT for $n$ qubits, then the non zero coefficients are
\begin{equation}
    \vec{h}_{n}=(x_n^{(0)}\ x_n^{(1)}\ x_n^{(2)}\ x_n^{(4)}\ \ldots \ x_n^{(2^{n-1})} )
\end{equation}
with $|x_n^{(1)}|<|x_n^{(2)}|<\ \ldots <|x_n^{(2^{n-1})}|<~|x_n^{(0)}|$. We keep $x_0$ and the largest $k_0$ values of the coefficients with $|k|_b=1$, i.e. $\vec{h}_{n}^{(k_0:n)}=~(x_n^{(0)}\ 0\ 0\ldots 0\ x_n^{(2^{n-k_0})}\ \ldots \ x_n^{(2^{n-1})})$. Next, we construct  the circuit to generate the state encoding these renormalized coefficients of the truncated series. Then, the fidelity of the resulting state,  $|\Phi^{(k_0:n)}_L\rangle_n$, with the exact state,  $|\Phi_L\rangle_n$, is given by
\begin{equation}\label{eq:fidelity}
  F=\frac{\frac{1}{4} \left(-1 + 2^{n}\right)^2 + \frac{1}{3} 2^{-2 + 2 n} \left(1 - 2^{-2k_0}\right)}{\frac{1}{4} \left(-1 + 2^{n}\right)^2 + \frac{1}{3} 2^{-2 + 2 n} \left(1 - 2^{-2n}\right)}. 
\end{equation}
More details about how to derive this expression are given in the Appendix A. \\ 

Note that while the structure of the circuit is the same, see Fig.~\ref{Fig:circuit_linear}, the angles have changed its value to $\theta^{k_0}_k=~\arcsin (2 x^{(2^k)}_{n}/\tilde{C}^{(k_0:n)}_n G^{(k_0:n)}_n(k))$, with $\tilde{C}^{(k_0:n)}_n$ the new normalization factor and $G^{(k_0:n)}_n(k)=~\prod_{i=k+1}^{n-1} \cos(\theta^{k_0}_i/2)$ if $n-k_0\leq k<n-1$ and $G^{(k_0:n)}_n(k)=1$ if $k=n-1$.

Now, assuming an infidelity $\epsilon=1-F$, it is possible to obtain the expression 
\begin{equation}
k_0=-\frac{1}{2}\log_2\bigg[ \frac{1}{2^{2n}} + \epsilon \bigg(4+\frac{2(1-3\cdot2^{n})}{2^{2n}}\bigg)\bigg]
\label{eq:ko_eps}
\end{equation}  
which establishes the trade off between infidelity and truncation. In the asymptotic limit $n\rightarrow \infty$, we obtain $k_0=\frac{1}{2}\log_2\left[ 1/(4\epsilon) \right].$
\\
\\

Additionally to this analysis, we also study the deviation of the amplitudes of the approximated state, $ \ket{\Phi^{(k_0:n)}_{L}}_n$ with respect to the ideal state, $\ket{\Phi_L}_n$. We now focus on describing the state $ \ket{\Phi_{L}^{(k_0:n)}}_n$ that comes from the truncation of the series, denoted as $\vec{h}_{n}^{(k_0:n)}=~(x_n^{(0)}\ 0\ 0\ldots 0\ x_n^{(2^{n-k_0})}\ \ldots \ x_n^{(2^{n}-1)})$. We compare the $k_0$ qubit state obtained from  $\vec{h}_{k_0}^{(k_0:n)}=~(x_n^{(0)}\ x_n^{(2^{n-k_0})}\ \ldots \ x_n^{(2^{n-1})})$ with the state produced by the series corresponding on doing the DHWT to the linear function encoded on $k_0$ qubits given by $\vec{h}_{k_0}= (x_{k_0}\ x_{k_0}^{(0)}\ x_{k_0}^{(2^{0})}\ \ldots \ x_{k_0}^{(2^{k_0-1})})$. From this comparison we can obtain that $$\alpha:=x_{n}^{(2^{n-k_0+j})}/x_{k_0}^{(2^{j})}= 2^{3/2(n-k_0)}$$ $$ \forall \ \ j=0,\ ...,\ k_0-1. $$

Therefore we can write 
\small
\begin{equation*}
    \vec{h}_{k_0}^{(k_0:n)}=\alpha (x_{k_0}^{(0)}\  x_{k_0}^{(2^{0})}\ \ldots \ x_{k_0}^{(2^{k_0-1})}) 
\end{equation*}

\begin{equation}
     + \beta (2_{k_0/2}^{(0)}\ 0\ \ldots\  0)
\end{equation} \normalsize
with $\beta=2^{(n-k_0)/2-1}(2^{n-k_0}-1) $ the offset that will induce the maximum deviation in the amplitude between both the ideal and approximated quantum state.
Thus, we can conclude that when truncating the sate, we are loading the $k_0$-qubit state $\ket{\Phi^{(k_0:n)}_{L,k_0}}=~\frac{1}{C^{(k_0:n)}_{k_0}}\sum_{j=0}^{2^{k_0}-1}(\alpha j + ~\beta) \ket{j}$ on the most significant $k_0$ qubits, with $C^{(k_0:n)}_{k_0}$ the corresponding normalization factor. When adding the remaining $n-k_0$ qubits, this will lead to an state with a degeneracy $2^{n-k_0}$ on every of this amplitudes (graphically a step wise function)

\small
\begin{equation*}
 \ket{\Phi^{(k_0:n)}_L}_n=~\frac{1}{C^{(k_0:n)}_n}\sum_{j=0}^{2^{k_0}-1}\sum_{l=0}^{2^{n-k_0}-1} (\alpha j + \beta) \ket{j} \ket{l}
\end{equation*}
\normalsize
with \small
\begin{equation*}
C^{(k_0:n)}_n=2^{n/2}\bigg[\frac{\alpha^2}{6} (2^{k_0+1}-1)(2^{k_0}-1) \\ 
\end{equation*}
\begin{equation}
    + \alpha\beta  (2^{k_0}-1) + \beta^2\bigg]^{1/2}.
\end{equation} \normalsize
Finally, merging both sums we finally get 
\small
\begin{equation}
 \ket{\Phi^{(k_0:n)}_L}_n=~\frac{1}{C^{(k_0:n)}_n}\sum_{j=0}^{2^{n}-1}(\alpha \lfloor j /2^{n-k_0}\rfloor + \beta) \ket{j}.
 \label{eq:k0state}
\end{equation}
From these expression we can define the deviation of the amplitude of $\ket{j}$ as 

\begin{equation}\label{error_amp}
 \delta^{(2)}_j:=\Bigg| \frac{1}{C^{(k_0:n)}_n}(\alpha \lfloor j /2^{n-k_0}\rfloor + \beta) -  \frac{j}{C_n}\Bigg|\leq \beta/C^{(k_0:n)}_n
\end{equation}
\normalsize

In the asymptotic limit the upper bound of $\delta_j^{(2)}$  given by $\beta/C^{(k_0:n)}_n$ decays as $\mathcal{O}(\frac{1}{\sqrt{2^n}})$ with a dominant coefficient given by \begin{equation} \frac{1}{2^{3k_0/2+1}\left(\frac{(2^{k_0}-1)(2^{k_0}-1)}{6\cdot 2^{3k_0}}+ \frac{2^{k_0}-1}{2^{3k_0+1}}+ \frac{1}{2^{3k_0+2}}\right)^{1/2}},\end{equation} which converges to 0 when $k_0$ tends to infinity as $\mathcal{O}(\frac{1}{2^{k_0}})$. We depict the complexities of loading the exact and approximated versions of linear function either via MPS or DWHT in Tab. \ref{tab:complexities_linear}.   \\

\begin{table}[b!]
\centering
\begin{tabular}{c||c||c}
\hline
\hline
  \diagbox[]{qubits}{method}& \textbf{Exact } & \textbf{Aprx} \\
\hline
\hline
$k_0$  & $\mathcal{O}(k_0)$  & - \\
$n$  & $\mathcal{O}(n)$  & $\mathcal{O}(k_0(\epsilon)$) \\

\hline
\end{tabular}
\caption{Complexities for loading the linear function into a different number of qubits $k_0$ and $n$, with $k_0<n$, either in an exact or approximated form. $k_0(\epsilon)$ corresponds to Eq. (\ref{eq:ko_eps}).  }
\label{tab:complexities_linear}
\end{table}

Alternative to this set up, one can also consider a MPS based methodology to load the linear function, which in the exact case scales linearly on the number of qubits, see Section \ref{From MPS to circuit}.  Thus, from now on, we will assume that we have access to a state preparation oracle of the exact linear function on $k_0$ qubits denoted as $U_{L,k_0}$, for $1\leq k_0\leq n$, along with its adjoint and their controlled variants. This allows us to load the linear function either in an exact manner with a circuit depth that scales $\mathcal{O}(k_0)$. Note that in the worst case scenario, the controlled version can be achieved by controlling every gate of the oracle, which keeps the complexity invariant.

\begin{table*}
\centering
\begin{tabular}{c||c||c}
\hline
\hline
\textbf{Parameter} & \textbf{Description} & \textbf{Value} \\
\hline
\hline
SQG time & Single qubit gate time (ns) & 35 \\
CX time & CX gate time (ns) & 540 \\
rD & Deviation ratio for the single qubit gates & 2,457E-04 \\
Pbf & Bit-flip error during the rz gate & 2,457E-04 \\
CNOTerror & Deviation ratio for CX gate & 8,328E-03 \\
pmeas & Readout error & 2,23E-01 \\
pth & Thermal population of the ground state & 0.01 \\
T1 & Decoherence time (us) & 214.84 \\
T2 & Dephasing time (us) &  214.84 \\
\hline
\end{tabular}
\caption{Noise parameters description and their value. We have estimated the numerical values from the calibration data provided for the IBM device `ibm\_jakarta'.}
\label{tab:features}
\end{table*}

\subsection{Polynomial transformation of amplitudes via quantum singular value transformation}
Once we have introduced the quantum circuit that loads the linear function into $k_0$ qubits via the unitary operator $U_{L,k_0}$ with a complexity of $\mathcal{O}(k_0)$, our method uses the quantum singular value transformation (QSVT) \cite{PhysRevX.6.041067, PhysRevLett.118.010501, Low2019hamiltonian, Gily_n_2019, tang2023cs, PhysRevA.103.042419} to achieve the polynomial transformation of the amplitudes. One also could consider the unitary to load the MPS of the linear function with $\chi=1$ on $k_0$ qubits, although the final results are considerable worst as we show later.

In this work we follow the procedure detailed in Ref.~\cite{Guo_2021} and in its subsequent exponential improvement in Ref.\cite{rattew2023nonlinear}, which provides a generalization that presents a diagonal block encoding and introduces the importance sampling. The remarkable insight of these works lies in the authors' demonstration of how it is possible to use the QSVT polynomial transformation to explicitly construct quantum circuits that apply the polynomial transformation to the amplitudes of any quantum state of interest, provided that the encoding unitary is known. 

The steps needed to achieve the state amplitude encoding of the polynomial are: the block encoding of the amplitudes, the polynomial transformation of the block encoding and the quantum amplitude amplification algorithm.

\subsubsection{Block encoding} 

The first step is the block encoding of the amplitudes of the linear function given the unitary $U_{L, k_0}$ that prepares the state that represents the linear function.

\begin{defn}
\cite{chakraborty_et_al:LIPIcs:2019:10609, PhysRevX.6.041067, PhysRevLett.118.010501, Low2019hamiltonian, Gily_n_2019, tang2023cs, PhysRevA.103.042419, Li_2023} Let be $A$ a $k_0$ qubit operator, $\alpha, \varepsilon \in~{\rm I\!R^+}$ and $a \in {\rm I\!N}$. We say that the $(a+k_0)$-qubit unitary $U_A$ is an $(\alpha, \text{a}, \varepsilon)$-block encoding of A if
\begin{equation}
    \| A - \alpha (\langle 0|^{\otimes a} \otimes I) U_A (|0\rangle^{\otimes a } \otimes I) \|_2 \leq \varepsilon.
\end{equation}
where $\alpha$ is the proportionality constant, $a$ the number of ancilla qubits required and $\epsilon$ the error achieved measured with  $||A||_2=\sigma_{\max}(A)$.

\end{defn}

According to Theorem 2 in Ref.~\cite{rattew2023nonlinear}, given a $k_0$-qubit state $\ket{\psi}=\sum_{j=0}^{2^{k_0}-1}\psi_j \ket{j}$ with real amplitudes loaded by a unitary $U$, i.e. $U\ket{0}^{\otimes n}=~\sum_{j=0}^{2^{k_0}-1}\psi_j \ket{j}$, $\psi_j \in \mathbb{R}$, it is possible to prepare a (1, ${k_0}+3$, 0) block encoding $U_A$ of $A=~\sum_{j=0}^{2^{k_0}-1}\psi_j \ket{j}\bra{j}$ with $\mathcal{O}({k_0})$ circuit depth and $\mathcal{O}(1)$ queries to a controlled version of $U$. Note that if the amplitudes are real the number of ancillary qubits required for the block encoding is ${k_0}+2$.

Here we use this result to create from $U_{L,k_0}$ the unitary $U_{A_{L,k_0}}$ which is a (1, $k_0+2$, 0) block encoding of the Hermitian operator $A_{L,k_0}= ~\frac{1}{C_{k_0}}\sum_{j=0}^{2^{k_0}-1}j \ket{j}\bra{j}$ that encodes the amplitudes of $\ket{\Phi_L}_{k_0}$ with $\mathcal{O}(k_0)$ circuit depth, see Appendix C for further details.

An alternative $(1,1,0)$-block encoding $U_B$ resulting by following the idea of Ref.~\cite{mcardle2022quantum} could be implemented by applying the unitary dilation technique \cite{constantinescu1996schur} to $B=~\frac{1}{2^k_0-1}\sum_{j=0}^{2^k_0-1}j\ket{j}\bra{j}$, given that $\|B\|\leq 1$. This operation would require an efficient simulation of the Hamiltonian $H=~\text{arccos}(B)$ \cite{Wang_2020}. Discussions of more possible block encoding for the linear function are shown in Appendix C.

\subsubsection{Polynomial transformation of the block encoding}

Once that we have obtained the amplitude block encoding, we show how to implement polynomial transformations of complex amplitudes via the Quantum Singular Value Transformation (QSVT) \cite{Gily_n_2019, Guo_2021, rattew2023nonlinear, mcardle2022quantum, Li_2023}. The construction and efficiency of explicit quantum circuits for applying the polynomial transformation to amplitudes of a $k_0$-qubit quantum state rely on the complexity of implementing the block encoding of the amplitudes, denoted as $U_{L,k_0}$. It is crucial that the complexity of $U_{L,k_0}$ remains of the order $\mathcal{O}(k_0)$ to prevent it from dominating the overall algorithmic cost.

\begin{lem}(Lemma 7 in Ref. \cite{rattew2023nonlinear}) Let $\gamma>0$. Given a (1, $k_0+2$, 0) block-encoding 
$U_A$ of $A=\sum_{j=0}^{2^{k_0}-1}\psi_j \ket{j}\bra{j}$ and $P(x)$ a polynomial with complex coefficients of degree $d$, such that $|P(x)|\leq~1/4\ \forall x$. Then, it is possible to obtain a (1, $k_0+5$, $\delta$) block encoding $U_{P(A)}$ of $P(A)$ with $\mathcal{O}(d k_0)$ circuit depth and $\mathcal{O}(d )$ calls to the controlled version of $U_A$ and $U^\dagger_A$. The circuit (rotation angles) can be computed with classical time complexity of $\mathcal{O}(poly(d,\log(1/\gamma))$ \cite{chao2020finding, Haah2019product, PhysRevA.103.042419}.
\label{lem2}
\end{lem}

By applying Lemma \ref{lem2} with a polynomial $P(x)$ to $U_{L,k_0}$ one obtains a block encoding $U_{P,k_0}$ of $P(A_{L,k_0})$. Additionally, note that due to the normalization, the values of the amplitudes of the state encoding the linear function have been renormalized to the interval $[0, (2^{k_0}-1)/\sqrt{(2^{{k_0}+1}-1)(2^{k_0}-1)2^{k_0}/6}]$. Therefore, to standardize the application of polynomials to the interval [0,1], we must compose the polynomial with the rescaling transformation, which keeps the degree of the polynomial invariant. For simplicity's sake, we will ignore this transformation from now on.

\subsubsection{Amplitude amplification}

We now apply the unitary $U_{P(A)}$ to the initial state $\ket{+}^{\otimes k_0}\ket{0}^{\otimes k_0 +5} $. This results into a quantum state $\frac{1}{4 \sqrt{2^{k_0}} ||P||_\infty} \sum_j P(   j /C_{k_0}) |j\rangle\ket{0}^{\otimes k_0 +5} + \dots$. 
The success probability of measuring \small $\ket{\Phi_P}_{k_0}=~\frac{1}{4 \sqrt{2^{k_0}} ||P||_\infty} \sum_j P(  j /C_{k_0}) |j\rangle\ket{0}^{\otimes k_0 +5}$ \normalsize is given by $0.0625 \mathcal{F}_P^2$, with  $\mathcal{F}_P$ being the filling ratio defined as \begin{equation}
    \mathcal{F}=\frac{||P||_2}{\sqrt{2^{k_0}}||P||_\infty},
\end{equation} and $||P||_\infty=\text{max}_{x \in [0,1]}|P(x)|$.
Therefore the quantum amplitude amplification algorithm would require $\mathcal{O}(4/\mathcal{F})$ rounds of the oracle that prepares the state to boost the success probability to 1 \cite{mcardle2022quantum}.

Note that the complexity of this protocol depends on the polynomial encoding that we aim to achieve. In this sense, some particular transformations will head to an efficient circuit, while others will introduce an exponential consumption of resources, depending on the filling ratio $\mathcal{F}$.

Once that we have obtained the state $\ket{\Phi_P}_{k_0}$ we can tensor it with the state $\ket{+}^{n-k_0}$ to obtain a quantum state $\ket{\Phi_P^{(k_0:n)}}_{n}$=$\ket{\Phi_P}_{k_0}\otimes  \ket{+}^{n-k_0}$ that approximately encodes the polynomial $P$ on $n$ qubits.

Before concluding this section, we would like to remark that for the particular case the polynomial transformation satisfies $P(0)=0$, one can use the importance sampling technique recently presented in Ref. \cite{rattew2023nonlinear} (Theorem 3) to achieve an exponential enhancement in the overall complexity do to the improvement in the step that encodes the transformed amplitudes into the final state.

\begin{table*}
    \centering
    \begin{tabular}{c||c||c||c}
    \hline
    \hline
    \textbf{Method} & \textbf{$\mathcal{F}$} & \textbf{$L_2$ norm} & \textbf{Fidelity} \\
    \hline
      \hline
    DHWT $(k_0=1)$ + QSVT   & 0.8164 & 0.1506 & 0.0753 \\
    DHWT $(k_0=2)$ + QSVT  & 0.6864 & 0.1486 & 0.0858 \\
    DHWT $(k_0=3)$ + QSVT  & 0.6331 & 0.0828 & 0.6095 \\
    Direct Pol MPS $(\chi=1)$  & - & 0.0752 & 0.6712 \\
    Lin MPS $(\chi=1)$ + QSVT  & - & 0.0701 & 0.7099 \\
    DHWT $(k_0=4)$ + QSVT  & 0.6190 & 0.0370 & 0.9144 \\
    Direct Pol MPS $(\chi=2)$  & - & 0.0252 & 0.9597 \\
    DHWT $(k_0=5)$ + QSVT  & 0.6183 & 0.0141 & 0.9874 \\
    Direct Pol MPS $(\chi=3)$  & - & 0.0049 & 0.9985 \\
    DHWT $(k_0=6)$ + QSVT  & 0.6184 & 0 & 1 \\
    Direct Pol MPS $(\chi=4)$  & - & 0 & 1 \\
    \hline
  \end{tabular}
    \caption{Error in $L_2$ Norm (fourth column) and Fidelities (fifth column) for Different Loading Methods of $P(x)=~\frac{1}{C_p}(x-1/(2^n-1))(x-20/(2^n-1))(x-50/(2^n-1))(x-60/(2^n-1))$ defined on $[0,1]$. Methods are arranged in ascending order based on fidelity of the polynomial encoding (fifth column). We also show the fidelity of loading the linear functions for those 2 steps methods (second column) and the filling ratio $\mathcal{F}$ (third column) for the different $k_0$ values.}
    \label{tab:fidelities_poly}
\end{table*}

\subsection{Amplitude deviation in the polynomial encoding}
 
 Finally, once the polynomial transformation has been achieved, we can analyze how the error coming from the approximation of 
 loading the exact $k_0$ qubit polynomial into $n$ qubits. In order to set a proper comparison with Ref. \cite{mcardle2022quantum, Li_2023}, we compute the error in the same way,

\begin{equation}
\delta=\text{max}_x\Bigg\Vert \frac{P_\text{exact}(x)}{||P_\text{exact}(x)||_\infty}- \frac{P_\text{aprx}(x)}{||P_\text{aprx}(x)||_\infty}\Bigg\Vert_\infty
\end{equation}
for our particular case we assume $||P_\text{exact}(x)||_\infty=~M$ and therefore 

\begin{equation*}
    \delta=\text{max}_x \frac{1}{M}\Bigg\Vert P_\text{exact}(x)- P_\text{exact}(x\pm \frac{2^{n-k_0}-1}{2^n-1}) \Bigg\Vert_\infty
\end{equation*}

\begin{equation}
    \leq \text{max}_x \frac{|P'(x)| \frac{2^{n-k_0}-1}{2^n-1}}{M}
\end{equation}

If we now assume $P(x)=\sum^d_{k=0}c_k x^k$, then
\begin{equation}
    \delta_\infty \leq  \frac{\frac{d^2-d}{2} \frac{2^{n-k_0}-1}{2^n-1}}{M}D
\end{equation}
with $D=\text{max}_k\{|c_k|\}$. In the asymptotic limit this upper bound decays as $\mathcal{O}(\frac{1}{2^{k_0}})$.

Note that one might think that it can be much cheaper to prepare an appropriate product state and then try to achieve an approximation to the desired polynomial encoding by applying a polynomial function to the product state via QSVT. However, since the complexity of the block encoding is not dominated by the oracle that loads the linear function, a reduction in complexity would not be achieved. Additionally, due to the approximation of MPS,this leads to a non-uniform grid discretization would result in a higher overall error when transforming the amplitudes.

\section{Numerical Results}
\label{num_results}
Having established the theoretical framework, in this section we present numerical simulations comparing the methods outlined in this paper. The primary objective is to provide empirical support for our analytical findings.

\subsection{Linear function}
We begin our analysis by evaluating the performance of the exact and approximated loading of the linear function and its its robustness in terms of fidelity against noise. 

When performing noisy simulations, the quantum circuit is transpiled to a native set of gates, including \textit{CNOT, Id, Rz $(\theta)$, X}, and \textit{Sx}. We consider various noise quantum channels, namely, bit-flip (Pbf), amplitude-damping (T1), dephasing (T2), gate errors (rD, CNOT error), and measurement error (pmeas). The specific noise parameters used in the simulations are summarized in Tab.~\ref{tab:features}.

The analysis of loading the linear function is depicted in Fig.~\ref{linear_analysis}, where we have considered $n=6$ qubits. In  Fig.~\ref{linear_analysis} (i) we have depicted a comparison of the ideal case $\chi=1$ for the MPS technique, the noisy and ideal cases with $k_0=3$ for the DHWT, and the exact encoding. In Fig.~\ref{linear_analysis} (ii) and (iii), we observe a trade-off between the truncation error and the experimental error for different values of $k_0$ when using the DHWT technique. The two figures of merit considered are the fidelity and the \(L_2\) norm, which is defined as \(\lVert \Vec{v} \rVert_2 = \left( \frac{1}{2^n} \sum \lvert v_i \rvert^2 \right)^{1/2}\), where \(\Vec{v}\) corresponds to the difference between the approximation and the ideal quantum state. We observe that the highest fidelity and the smallest $L_2$ norm error are achieved at truncation levels $k_0=2$ and $k_0=3$. In the comparison, we also include the ideal MPS with $\chi=1$.

\begin{figure*}[t!]
\centering
\includegraphics[width=.95\textwidth]{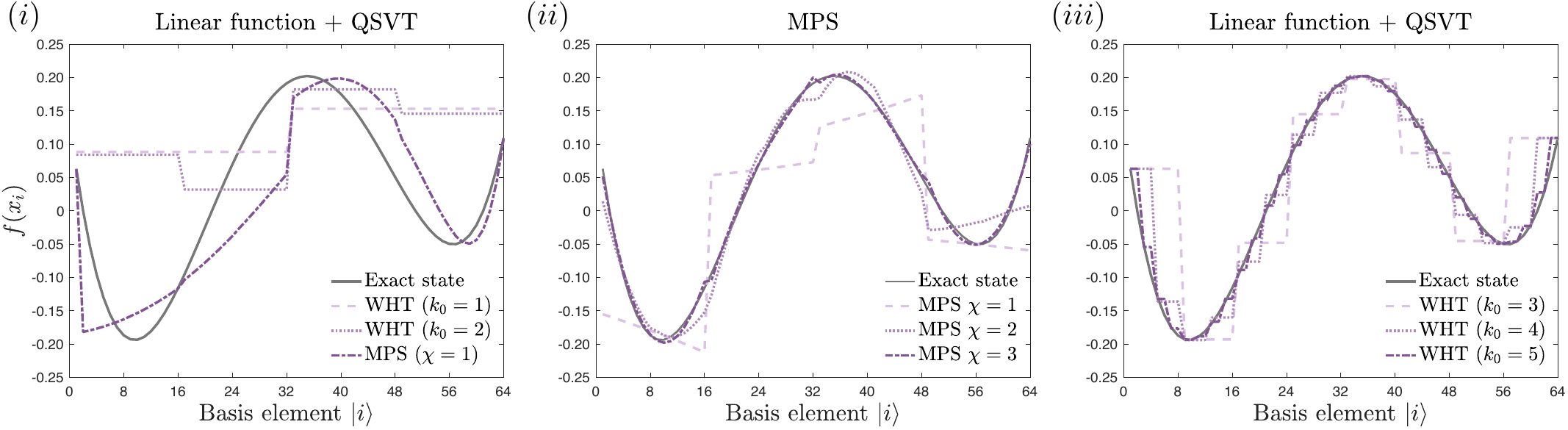}
\caption{Different methods to load the polynomial function  $P(x)=\frac{1}{C_p}(x-1/(2^n-1))(x-20/(2^n-1))(x-50/(2^n-1))(x-60/(2^n-1))$ for $n=6$ qubits using different methods. For this particular example the filling ratio for $k_0=6$ is $\mathcal{F}= 0.6184$. The worst and best implementation methods in which the linear function is loaded first, followed by the application of QSVT are displayed in $(i)$ and $(iii)$, respectively. In $(ii)$, results for the loading using MPS are shown, achieving perfect loading with $\chi=4$, although theory predicts $\chi=5$ as the upper bound. The $L_2$ norm and fidelities for each case are displayed in Tab. \ref{tab:fidelities_poly}.}
\label{Fig:example_polynomial_MPS}
\end{figure*}

\subsection{Example of polynomial function}
We present a second analysis where we consider the encoding of a polynomial function. In Fig.~\ref{Fig:example_polynomial_MPS}, we present the loading of the polynomial function \small $$ P(x)=\frac{1}{C_p}(x-1/(2^n-1))(x-20/(2^n-1)) $$ $$(x-50/(2^n-1))(x-60/(2^n-1)),$$\normalsize with $C_p$ the normalization factor,  by using $n=6$ qubits and the two methods studied in this paper. Our approach involves two distinct loading strategies: first, loading the linear function and then applying the polynomial transformation with the QSVT method, ensuring no induced errors (Figs.~\ref{Fig:example_polynomial_MPS} $(i)$ and $(ii)$), and second, using the matrix product state (MPS) representation of the polynomial itself (Fig.~\ref{Fig:example_polynomial_MPS} $(iii)$). For additional examples see Appendix D.

In order to load the linear function, we utilize either the DHWT method introduced in this work for different truncation values $k_0$, or the MPS approach with $\chi=1$. In Tab.~\ref{tab:fidelities_poly} we show the $L_2$ norm and fidelity of the final state with respect to the exact state for each methodology. We observe that for $k_0\geq 5$, the value of the fidelity achieved by the combined protocol is better than the direct polynomial MPS technique with $\chi=2$.

Additionally, the fidelities resulting from combining the approximate loading of the linear function with the QSVT to perform the polynomial transformation are presented. Remarkably, the loading of the linear function via a $\chi=1$ MPS achieves a fidelity of $F=0.9885$. This suggests that the quantum state of the linear function is nearly a product state. To gain deeper insights, we performed an analysis of the single-qubit rotations that generate this approximate state. We fitted the angles of the rotations to an analytical expression and leveraging this fitting information, we trained a variational circuit aimed at preparing a product state that optimizes the fidelity with respect to the exact linear function. It is crucial to emphasize that the effectiveness of the fitting is not  guaranteed across different number of qubits, hence we undertook this approach as a validation method. For additional details, please refer to the Appendix B.

\begin{table*}
\centering
\begin{tabular}{c||c||c||c}
\hline
\hline
& \textbf{Sine encoding \cite{mcardle2022quantum, Li_2023}} & \textbf{This work QSVT} & \textbf{MPS} \\
\hline
\hline
Block encoding  & 1+2 \text{ Ancillas} & (n+2)+2 \text{ Ancillas}& n/a \\
 & $\mathcal{O}(n)$ \text{ Gates} & $\mathcal{O}(n)$ \text{ Gates} \cite{Guo_2021, rattew2023nonlinear} &  \\
 \hline
Exact block encoding   & n/a & $\mathcal{O}(n d)$ & $\mathcal{O}(n2^{d})$ \\ 
poly degree $d$  &  &  &    \\ \hline
Approximated block   & $\mathcal{O}(n\ d(\delta_\infty))$ \cite{mcardle2022quantum} & $\mathcal{O}(d\ k_0(\delta_\infty))$  & $\mathcal{O}(n)\:(\chi=2)$   \\
encoding &  $d(\delta_\infty)\sim\frac{\log(B/\delta_\infty)}{\nu}$ & $k_0(\delta_\infty)\sim \log\left(\frac{D(d^2-d)}{2\delta_\infty M}\right)$  & Non controllable    \\
poly degree $d$ & $\nu=1-\sin((2^n-1)/2^n)$  &  & error \\
\hline 
 Amplitude Amplification & $\mathcal{O}(n\ d(\delta_\infty)/\mathcal{F})$   &  $\mathcal{O}( k_0(\delta_\infty)d /\mathcal{F})$ & n/a \\
\hline
\end{tabular}
\caption{Comparison between the methodology proposed in Refs. \cite{mcardle2022quantum, Li_2023} and the current work, with $||P_\text{exact}(x)||_\infty=~M$, $\mathcal{F}$ the filling ratio, $D=\text{max}_k\{|c_k|\}$ and   $\delta_\infty=||P(j)/||P(j)||_\infty-\tilde{P}(j)/||\tilde{P}(j)||_\infty||_\infty$, where $P=\sum_{k=0}^d c_k x^k $ is the exact polynomial and $\tilde{P}$ its approximation, and $B\geq \sum_{k}|c_k|$. We neglect amplitude amplification and precomputational errors and complexities. In both cases where QSVT is used we have assumed no error in the classical preporcesing of the angles needed to implement the polynomial transformation.}
\label{tab:comparison}
\end{table*}

\section{Comparison with other methods}

The first statement we would like to highlight is that we have put large part of our efforts in achieving an approximate loading of functions as simple as the linear function, by introducing a controllable error that reduces the depth of an already efficient circuit. As far as our knowledge is, there is no previous result in the literature that can do the same task with a comparable performance. That being said, we proceed to compare our method with similar results.

The QSVT enables the application of polynomial transformations to the amplitudes of a quantum state. However, when utilizing this technique to load a polynomial function encoded in the quantum state's amplitudes, the efficient loading of the linear function is crucial. Previously in literature, authors in Ref. \cite{mcardle2022quantum, Li_2023} explored the possibility of applying the QSVT to block encoding of the sine function.
Our method posses the same advantages of using QSVT that they mention: `we avoid discretizing the values the function can take, providing instead a continuous approximation to the function. Our method is straightforward and versatile, as the same circuit template can be used for a wide range of functions. In contrast, our method avoids the error that propagates into the final loaded state due to the polynomial approximation of the $\arcsin$ function by efficiently loading the block encoding of linear function, and the subsequent polynomial transformations applied to load the desired polynomials into the amplitudes. In this context, our approach focuses on implementing the block encoding of the linear function rather than the sinusoidal function. To achieve this, we have proposed a method based on the Hadamard-Walsh Transform. The cost of this replacement in the block encoding can me mainly expressed in terms of adding $n+1$ extra ancillary qubits to the circuit, see Tab. \ref{tab:comparison} for comparison. From this table we can observe that
or the encoding of a polynomial without error, our algorithm scales with the same complexity as \cite{mcardle2022quantum, Li_2023} with the error resulting from the arcsin approximation. In order to explore the polynomial degree overhead needed in the methodology presented in Ref \cite{mcardle2022quantum} to encode a polynomial into amplitudes of a quantum state we present Tab.~ \ref{tab:berta_fidelities}. Additionally, our methodology allows the introduction of a controllable error that reduces the complexity of the exact case for the encoding of polynomial functions. In the general case of loading an arbitrary function the trade off between the additional resources of our protocol versus the approximation of arcsin should be taken into account. For instance,  \cite{mcardle2022quantum, Li_2023} are more efficient for loading trigonometric polynomials. 

\begin{table}[b!]
    \centering
    \begin{tabular}{c||c}
    \hline
    \hline
    \textbf{Degree $d$} & \textbf{Fidelity} \\
    \hline
      \hline
    $4$ & 0.0736 \\
    $8$ & 0.4969 \\
    $10$ & 0.8477 \\
    $12$ & 0.9672 \\
    $14$ & 0.9931 \\
    $16$ & 0.9985 \\
    $18$ & 0.9997 \\
    $20$ & 0.9999 \\ 
    \hline
  \end{tabular}
    \caption{Fidelities corresponding to different degrees of the approximation polynomial employed to load \(P(x) =~ \frac{1}{C_p}(x - \frac{1}{2^n-1})(x - \frac{20}{2^n-1})(x - \frac{50}{2^n-1})(x - \frac{60}{2^n-1})\) using the method outlined in Ref.~\cite{mcardle2022quantum}.}
    \label{tab:berta_fidelities}
\end{table}
In addition to QSVT-based methods, there have been approaches utilizing matrix product states (MPS) for the loading of smooth differential real-valued functions (SDR) into quantum state amplitudes \cite{iaconis2023quantum,Grasedyck2010PolynomialAI,holmes2020efficient,Melnikov_2023}. Ref.~\cite{holmes2020efficient} shows that for such functions, favorable outcomes can be obtained by employing a fixed bond dimension of two, attributed to the logarithmic scaling of entanglement entropy in these cases \cite{holmes2020entanglement}. In this paper, we have explored the resource requirements of this approach and compare it to our linear function+QSVT approach, specially in the case of loading of the linear function, for which we have analytical expressions for fidelity and error propagation.

Considering a different perspective, it was recently proposed the use of the Hadamard-Walsh series \cite{zylberman2023efficient} for amplitude encoding. This proposal leverages the fact that for functions whose derivative is bounded, the error resulting from truncating their discrete Hadamard-Walsh Series is exponentially suppressed with the index of the truncation \cite{1672864}. The authors use Hamiltonian simulation techniques presented in \cite{Welch_2014} to achieve a Hadamard-Walsh approximated simulation of the unitary $U=e^{-i\hat{f}\epsilon_0}$ with error $\epsilon_1$, where $\hat{f}=~\sum_xf(x)\ket{x}\bra{x}$ is the operator corresponding to the target function to be encoded. Finally they use an ancillary qubit to generate the operator $-i(I-e^{-i\hat{f}\epsilon_0})$ acting on the state $\sum_x \ket{x}\ket{1}$, which approximates the target state to first order of $\epsilon_0$ and introduces a protocol conditioned to the probability of measuring the ancillary qubit in the state $\ket{1}$. Therefore the total protocol introduces two sources of error, $\epsilon_1 $ corresponding to the truncation of the Hadamard-Walsh series and $\epsilon_0$ corresponding to the Taylor expansion. This technique introduces an error for loading functions even for those cases in which $\epsilon_1=0$, as in the linear case. By contrast our subroutine that uses the DHWT leverages the sparsity of the series corresponding to polynomial functions to efficiently generate quantum circuits that encode the states directly into the amplitudes. Additionally, this state loading algorithm based on the Hadamard-Walsh transform is not as efficient as our methodology for the application of loading the linear function, which is an important factor when loading the linear function for the derivative pricing use case.

 Finally, we remark that our method is not limited by the condition of a bounded derivative of the target function, as is the case for the MPS \cite{holmes2020efficient} or the Hamiltonian simulation based on DHWT \cite{zylberman2023efficient}.

\section{Conclusions}
In this article, we have considered the problem of loading real polynomials into a quantum computer, with a particular focus on the encoding of the linear function. We have presented and reviewed two methodologies based on different approaches. The first method is based on matrix product states, and even though it offers competitive results for some particular cases \cite{holmes2020efficient}, it is difficult to precisely control the error the method incurs. The second algorithm introduced in this paper utilizes the discrete Hadamard-Walsh transform (DHWT) to achieve the block encoding of the amplitudes which is fed into the quantum singular value transformation (QSVT) in order to apply a polynomial transformation to the amplitudes.This technique is able to exactly encode polynomials with same complexity as previous works \cite{mcardle2022quantum, Li_2023} that incurred into an approximation error. Furthermore we have been able to reduce the complexity of the protocol via introducing a controllable error.

Using this technique based on the DHWT, the coefficients of the Hadamard-Walsh series of a given function are loaded into a quantum state and the inverse discrete Hadamard-Walsh transform (DHWT) is applied to achieve the amplitude encoding of the target function. This idea constitutes a novel and promising approach for functions whose Hadamard-Walsh series can be efficiently encoded, as for instance when it sparse \cite{PhysRevLett.129.230504} or efficiently truncated \cite{Welch_2014}.

We would like to remark that even though our work has been focused on encoding real polynomials, it could be easily extended to load polynomials valued on complex values, i.e. $P~: ~\mathbb{C} \to \mathbb{C}$ , multivariate polynomials or even non-linear functions approximated with polynomials \cite{mcardle2022quantum, Guo_2021, rattew2023nonlinear}. Future work will consider the feasibility of using our DHWT-based method to load highly discontinuous square wave-like functions.

\begin{acknowledgements}
We thank N. Guo and A. Rattew for the useful discussions regarding the amplitude transformations via the QSVT. We thank M. Cea-Fernandez for the discussions on the matrix product states. The 
authors acknowledge financial support from OpenSuperQ+100 (Grant No. 101113946) 
of the EU Flagship on Quantum Technologies, as well as from the EU FET-Open project 
EPIQUS (Grant No. 899368), also from Project Grant No. PID2021-125823NA-I00 595 and 
Spanish Ramón y Cajal Grant No. RYC-2020-030503-I funded by MCIN/AEI/10.13039/501100011033 and by “ERDF A way of making Europe” and “ERDF 
Invest in your Future,” this project has also received support from the Spanish Ministry for Digital Transformation and of Civil Service of the Spanish Government through the QUANTUM ENIA project call - Quantum Spain, and by the EU through the Recovery, Transformation and Resilience Plan – NextGenerationEU within the framework of the Digital Spain 2026 Agenda, we acknowledge funding from Basque Government through Grant No. IT1470-22 and the IKUR 
Strategy under the collaboration agreement between Ikerbasque Foundation and BCAM 
on behalf of the Department of Education of the Basque Government, as well as from 
and UPV/EHU Ph.D. Grant No. PIF20/276. PR acknowledges financial support from the CDTI within the Misiones 2021 program and the Ministry of Science and Innovation under the Recovery, Transformation and Resilience Plan—Next Generation EU under the project “CUCO: Quantum Computing and its Application to Strategic Industries”
\end{acknowledgements}

\newpage
\onecolumngrid
\onecolumngrid
\newpage
\section*{Appendix A Proofs}

In this appendix we derive the expression shown in Eq. \ref{eq:fidelity}. Let us first consider the exact state corresponding to the Walsh Hadamard series

$$\ket{\Tilde{\Phi}_L}=\frac{1}{\tilde{C}} \sum_{|k|_b\leq 1} x^{(n)}_k\ket{k}= x^{(0)}_n \ket{0...0} + x^{(2^0)}_{n} \ket{0...1} + ... + x^{(2^{n-1})}_{n} \ket{1...0}$$
with $\tilde{C}=2^{n/2} \sqrt{\left(\frac{2^n-1}{2}\right)^2+\frac{2^{2n}-1}{12} }$. If we truncate keeping the largest $k_0$ values with $|k|_b= 1$ and $x_0^{(n)}$, the renormalized state results 

$$\ket{\Tilde{\Phi}_L^{k_0}}=\frac{1}{\tilde{C}_{k_0}} \sum_{|k|_b\leq 1} x^{(k)}_n\ket{k}= x^{0}_n \ket{0......0} + x^{2^{n-k_0}}_{n} \ket{0...1 ... 0} + ... + x^{(2^{n-1})}_{n} \ket{1......0}$$

with $\tilde{C}_{k_0}=2^{n/2} \sqrt{\left(\frac{2^n-1}{2}\right)^2+\frac{1}{12}\left(2^{2n}-2^{2(n-k_0)}\right) }$.
When we calculate $$\bigg| \langle \Tilde{\Phi}_L |\Tilde{\Phi}_L \rangle_{k_0} \bigg|^2=\left(\frac{\tilde{C}_{k_0}}{\tilde{C}}\right)^2$$ and manipulating this expression, this lead to Eq. \ref{eq:fidelity}.

\section*{Appendix B Variational circuit for linear function}
In this article, we investigated the loading of a linear function using an MPS (Matrix Product State) with a bond dimension of $\chi=1$, which yielded remarkably high fidelities. These high fidelities suggest that the linear function closely resembles a product state, given that an MPS with $\chi=1$ can be efficiently constructed using single qubit rotations $R_y(\theta)$. Building on this insight, our focus in this section is to develop a variational algorithm to find the optimal angles that maximize the fidelity of our linear function approximation within a product state framework.

\begin{figure}[H]
    \centering
    \includegraphics[width=0.5\linewidth]{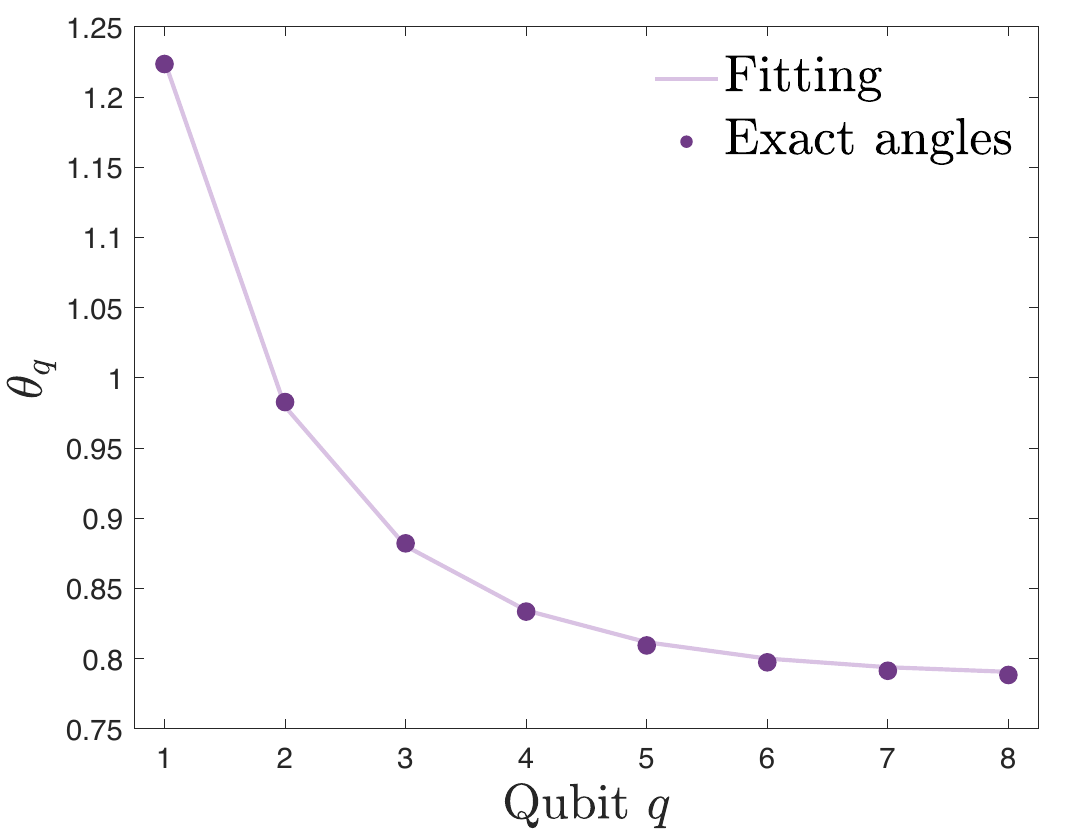} 
    \caption{Angles of the single qubit rotations to load the linear function through an MPS with $\chi=1$ for $n=8$ and its analytical fitting.}
    \label{fig:fitting}
\end{figure}

To initiate the process, we derived the angles from the MPS with bond dimension 1 for a system of $n=8$ qubits. Subsequently, we performed an analytical fitting for the angles, given by
\begin{equation}\label{fitting}
    \theta_q = \exp\left(\exp\left(-q^{0.9}/1.23\right)-0.24\right),
\end{equation}
where $q$ ranges from 1 to 8, representing each qubit. The fitting results are depicted in Fig.~\ref{fig:fitting}.
Moving on to our variational model, we utilized a system with $n=6$ qubits, aiming to maximize the fidelity of the linear function. We initialized the angles based on the fitting provided by Eq.~\ref{fitting}. To optimize this process, we employed gradient descent and employed the squared $L_2$ norm as the loss function to quantify the disparity between the exact linear function and the predicted state. For the $n=6$ qubit system, the fidelity between the exact linear function and the one obtained from the MPS with $\chi=1$ was $F_{\mathrm{MPS}}=0.9875747$. After the training, our variational model achieved an improved fidelity of $F_{\mathrm{var}}=0.9875750$.

\section*{Appendix C Block Encoding of Real-valued Statevector Amplitudes}
\label{block encoding}
In this paper we are primarily concerned applying the methods of Ref.~\cite{Guo_2021} to loading polynomial functions into quantum registers. In order to achieve this, we introduced an amplitude encoding of the linear function $U_L : |0\rangle_n \mapsto |\Phi_L \rangle \propto \sum_j j |j\rangle_n$. The unitary $U_L$ is used to construct a block-encoding $U_A$ of a matrix $A \propto \mathrm{diag}(0, 1,...,2^n-1)$ to which the either the QSVT can be applied to transform the diagonal entries by an arbitrary polynomial; alternatively products and linear combinations of these block-encodings can be used to more efficiently encode low degree polynomials. In this appendix, we consider how the methods of Ref.~\cite{Guo_2021} have been improved upon by Ref.~\cite{rattew2023nonlinear} in order to efficiently construct the block-encoding $U_A$.

When constructing the block-encoding $U_A$, we are concerned with the special case where the statevector amplitude loading unitary $U|0\rangle_n = \sum_j \psi_j |j\rangle_n$ has real-valued amplitudes $\psi_j \in \mathbb{R}$ i.e., $U=U_L$. In order to construct $U_A$, we first must define several unitary operators. Following the notation introduced in Ref.~\cite{rattew2023nonlinear}, we first define the operator $W_0$

\begin{equation}
   W_0 :=  (\mathcal{I}_n \otimes H \otimes \mathcal{I}_n) C U_C (\mathcal{I}_n \otimes H \otimes \mathcal{I}_n)
\end{equation}

\noindent where $U_C$ and $C$ are defined as
\begin{equation}
   U_C := (U \otimes |0\rangle \langle 0|_1 +  \mathcal{I}_n \otimes |1\rangle \langle 1|_1) \otimes \mathcal{I}_n
\end{equation}

\begin{equation}
   C :=  \mathcal{I}_n  \otimes  |0\rangle \langle 0|_1 \otimes \mathcal{I}_n + \sum_{k, j = 0}^{2^n - 1} |j \oplus k \rangle \langle j|_n \otimes  |1\rangle \langle 1|_1 \otimes |k\rangle \langle k|_n
\end{equation}

$U_C$ is implemented using a controlled version of the amplitude loading unitary $U$ and $C$ is "controlled-copy" circuit which can be implemented with a cascade of $n$ Toffoli gates, see Ref.~\cite{rattew2023nonlinear} for more details.

The $W_0$ operator sends states $|0\rangle_n |0\rangle_1 |k\rangle_n$ to $|\Phi_k^0\rangle = \frac{1}{2} ((|\psi\rangle_n + |k\rangle_n)|0\rangle_1 + (|\psi\rangle_n - |k\rangle_n)|1\rangle_1) |k\rangle_n $. The $W_0$ operator is used to construct another operator $G_0$ which is defined as  
\begin{equation}
   G_0 := W_0 ((\mathcal{I}_{n+1} - 2 |0\rangle \langle 0|_{n+1}) \otimes  \mathcal{I}_n) W_0^\dagger (\mathcal{I}_n \otimes Z \otimes \mathcal{I}_n)
\end{equation}

The operator $G_0$ has the important property that $|\Phi_k^0\rangle$ are its eigenvectors with eigenvalues $\psi_j \in \mathbb{R}$. $G_0$ and its inverse $G_0^\dagger$ can then be used to construct the desired $(1,n+2,0)$-block-encoding $U_A$ of $A = \mathrm{diag}(\psi_0, \psi_1,..., \psi_{2^n-1})$,
\begin{equation}
   U_A := (XZX \otimes \mathcal{I}_{2n+1})(H \otimes W_0^\dagger) ( |0\rangle \langle 0|_1 \otimes G_0 + |1\rangle \langle 1|_1 \otimes G_0^\dagger)(H \otimes W_0)
\end{equation}

As stated in Ref.~\cite{rattew2023nonlinear}, $U_A$ can be implemented with $O(n)$ circuit depth, 3 queries to a controlled-$U$ gate and 3 queries to an inverse controlled-$U$ gate.
\newpage
Additionally, one can also consider some alternative ways to achieve the block encoding of the linear function\\ \\ \begin{itemize}

         \item Using the rotation oracle acting on $\sum_{x} \ket{x}\ket{0}$ \cite{PhysRevLett.85.1334}: \\  $$\mathfrak{rot} \ket{x}\ket{0}\rightarrow  \ket{x}(\sin(\theta) \ket{0} + \cos(\theta)\ket{1}),$$ \\ 
         where $\theta$  is a $m$ bit approximation to \text{arcsin}($x$).
         To implement this procedure, the quantum computer would calculate $\theta$, store the value in $m$ ancillary registers, and use that register as the control for a sequence of rotation operations on the qubit. Therefore this step incurs into the use of coherent arithmetics.\\

         \item Using the comparing oracle acting on $\sum_{x} \ket{x}\ket{0}^n\ket{0}$. We firstly implement the Amplitude oracle that encodes the uniform superposition in a secondary ancilla register $\ket{x}$, $$\mathfrak{amp} \ket{x}\ket{0}^n\ket{0}\rightarrow \frac{1}{2^n}\sum_y\ket{x}\ket{y}\ket{0}.$$ This can be straightforwardly done by applying a layer of Hadamard gates to $n$ ancillary qubits.\\

         Next we use the comparison oracle \cite{PhysRevLett.122.020502}
         
         $$\mathfrak{comp}  \ket{x}\ket{y}\ket{0}\rightarrow   \left\{ \begin{array}{lcc}  \ket{x}\ket{y}\ket{0} & if & x < y \\ \\ \ket{x}\ket{y}\ket{1} & if & x \geq y \end{array} \right. $$ \\ By composing this two oracles, $\mathfrak{comp}\circ \mathfrak{amp}$, and apply them to the initial state $\sum_{x} \ket{x}\ket{0}^n\ket{0}$, and subsequently erasing the $n$-ancilla register, one obtain the state $\ket{x}(x/2^n \ket{0} + \sqrt{1-(x/2^n)^2}\ket{1})$.  The complexity of this protocol is $\mathcal{O}(n)$ and uses $n+1$ ancillas for the block encoding. Therefore, this is an equivalent methodology to achieve the same result that we have presented in the main text.
         \\ \\

           \item Using Hamiltonian simulation techniques applied to the unitary dilation of $B$. As we already mention in the main  text, by applying the unitary dilation technique \cite{constantinescu1996schur} to $B=~\frac{1}{2^n-1}\sum_{n=0}^{2^n-1}j\ket{j}\bra{j}$, given $\|B\|\leq 1$. This operation would require an efficient simulation of the Hamiltonian $H=~\hat{\sigma}^y\otimes\text{arccos}(B)$. This leads to the block encoding unitary 

           \begin{equation*}
            U=\begin{pmatrix}
            B & \sqrt{1-B^2} \\
            \sqrt{1-B^2} & -B
            \end{pmatrix}=
            \left(\hat{\sigma}^z\otimes \mathds{I} \right)\exp \left(i\hat{\sigma}^y\otimes H\right),
            \end{equation*}

            which is a $(1,1,0)$-block encoding, and could be efficiently implemented with the techniques shown in Ref. \cite{Wang_2020}.
         \\ \\
    
    \end{itemize}

\newpage
\section*{Appendix D Additional numerical examples}
Here we provide two more numerical experiments for $n=6$ qubits for two different polynomials.

\begin{table*}[h!]
    \centering
    \begin{tabular}{c||c|c|c||c|c|c}
    \hline
    \hline
      & \textbf{$P_1(x)$} & \textbf{$P_1(x)$} & \textbf{$P_1(x)$} & \textbf{$P_2(x)$} & \textbf{$P_2(x)$} & \textbf{$P_2(x)$}\\
     \textbf{Method} & \textbf{$L_2$ norm} & \textbf{Fidelity poly} & $\mathcal{F}$ & \textbf{$L_2$ norm} & \textbf{Fidelity poly} & $\mathcal{F}$\\
    \hline
      \hline
    Lin MPS $(\chi=1)$ + QSVT & 0.0790 & 0.6408 & - & 0.0570 & 0.8026 & -\\
    DHWT $(k_0=1)$ + QSVT & 0.1715 & 0.0034 & 0.7234 & 0.2357 & 0.6044 & 0.8539\\
    DHWT $(k_0=2)$ + QSVT & 0.1889 & 0.0202 & 0.5801 & 0.1522 & 0.0671 & 0.7884\\
    DHWT $(k_0=3)$ + QSVT & 0.1339 & 0.1818 & 0.5542 & 0.0700 & 0.7112 & 0.7010\\
    DHWT $(k_0=4)$ + QSVT & 0.0647 & 0.7498 & 0.5023 & 0.0304 & 0.9418 & 0.6677\\
    DHWT $(k_0=5)$ + QSVT & 0.0252 & 0.9598 & 0.4730 & 0.0115 & 0.9915 & 0.6539\\
    DHWT $(k_0=6)$ + QSVT & 0 & 1 & 0.4621 & 0 & 1 & 0.6551\\
    Direct Pol MPS $(\chi=1)$ & 0.0885 & 0.5619 & - & 0.0763 & 0.6622 & -\\
    Direct Pol MPS $(\chi=2)$ & 0.0291 & 0.9467 & - & 0.0105 & 0.9930 & -\\
    Direct Pol MPS $(\chi=3)$ & 0.0134 & 0.9885 & - & 0.0014 & 0.9999\\
    Direct Pol MPS $(\chi=4)$ & 0 & 1 & - & 0 & 1 & -\\
    \hline
  \end{tabular}
    \caption{Error in $L_2$ Norm and Fidelities for Different Loading Methods of two different polynomials $P_1(x)=~\frac{1}{C_p}(x-2/(2^n-1))(x-16/(2^n-1))(x-40/(2^n-1))(x-50/(2^n-1))(x-62/(2^n-1))$ and $P_1(x)=~\frac{1}{C_p}(x-2/(2^n-1))(x-32/(2^n-1))(x-60/(2^n-1))$. We also provide the filling ratios $\mathcal{F}$ for the different $k_0$ values}. 
    \label{tab:fidelities_poly}
\end{table*}

\begin{figure}[b!]
\centering
\includegraphics[width=1\textwidth]{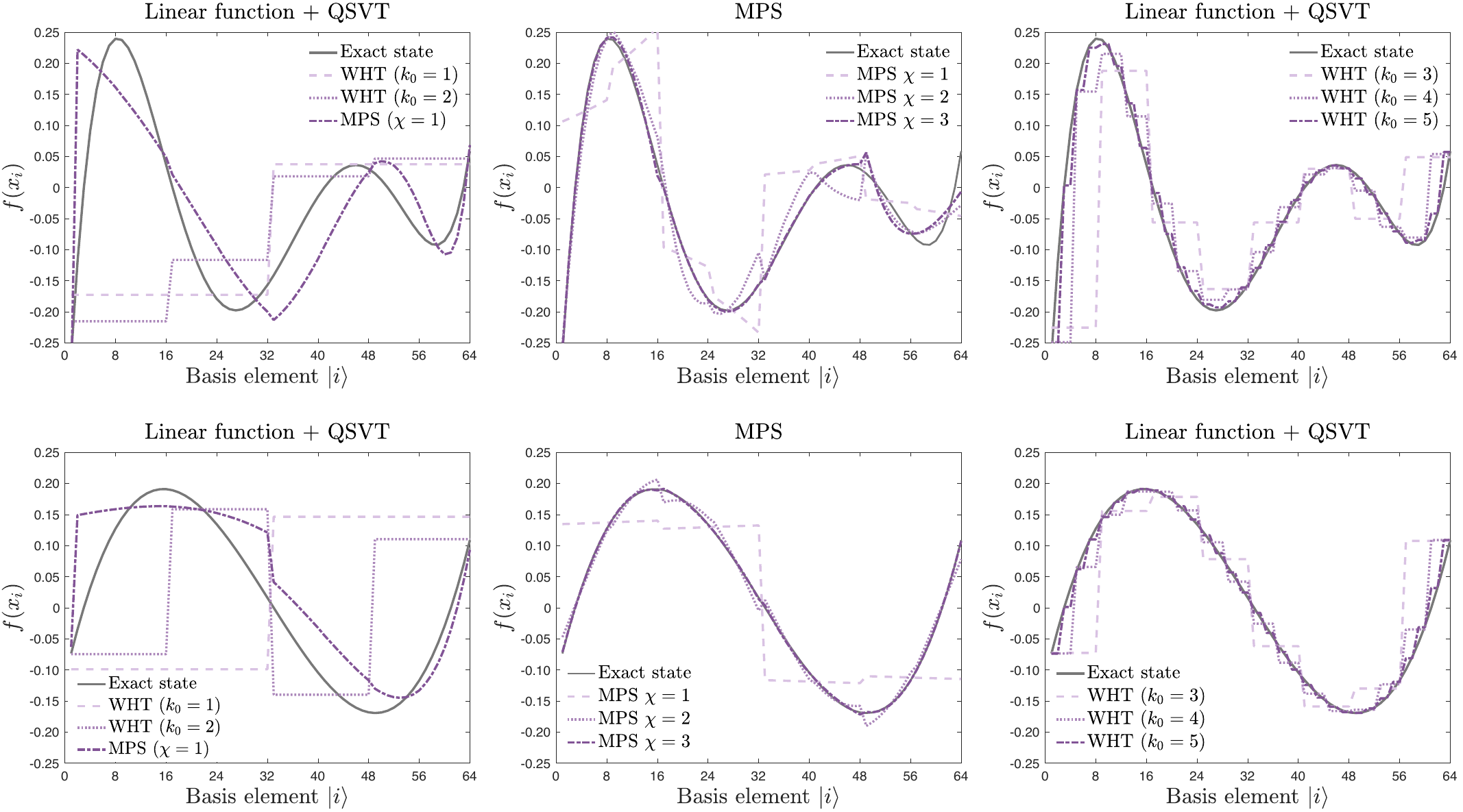}
\caption{Various approaches for loading polynomial functions \(P_1(x) = \frac{1}{C_p}(x-\frac{2}{2^n-1})(x-\frac{16}{2^n-1})(x-\frac{40}{2^n-1})(x-\frac{50}{2^n-1})(x-\frac{62}{2^n-1})\) in the upper portion of the figure and \(P_2(x) = \frac{1}{C_p}(x-\frac{2}{2^n-1})(x-\frac{32}{2^n-1})(x-\frac{60}{2^n-1})\) in the lower portion for \(n=6\) qubits. }
\end{figure}

\end{document}